\documentclass[12pt,preprint]{aastex}
\usepackage{epsfig}

\shorttitle{GRS~1915+105: QPO~Frequency Behavior}
\shortauthors{Mikles et al.}

\begin{document}

\title{Does Low Frequency X-ray QPO Behavior in GRS~1915+105 Influence Subsequent X-ray and Infrared Evolution?}
\author{Valerie J. Mikles}
\affil{Department of Astronomy, University of Florida, Gainesville, FL 32611}
\email{mikles@astro.ufl.edu}

\author{Stephen S. Eikenberry}
\affil{Department of Astronomy, University of Florida, Gainesville, FL 32611}
\email{eikenberry@astro.ufl.edu}

\and

\author{David M. Rothstein}
\affil{Department of Astronomy, Cornell University, Ithaca, NY 14853}
\email{droth@astro.cornell.edu}

\begin{abstract}
Using observations with the {\it Rossi X-ray Timing Explorer}, we examine the behavior of $2-10$~Hz quasi-periodic oscillations (QPOs) during spectrally-hard dips in the x-ray light curve of GRS~1915+105 that are accompanied by infrared flares. Of the twelve light-curves examined, nine are $\beta$-class and three are $\alpha$-class following the scheme of \citet{bell00}. In most cases, the QPO~frequency is most strongly correlated to the power law flux, which partially contradicts some earlier claims that the strongest correlation is between QPO~frequency and blackbody flux. Seven $\beta$-class curves are highly correlated to blackbody features. In several cases, the QPO evolution appears to decouple from the spectral evolution. We find that $\beta$-class light-curves with strong correlations can be distinguished from those without by their ``trigger spike'' morphology. We also show that the origin and strength of the subsequent infrared flare may be causally linked to the variations in QPO~frequency evolution and not solely tied to the onset of soft x-ray flaring behavior. We divide the twelve $\alpha$- and $\beta$-class light-curves into three groups based on the evolution of the QPO, the morphology of the trigger spike, and the infrared flare strength. An apparent crossover case leads us to conclude that these groups are not unique modes but represent part of a continuum of accretion behaviors.  We believe the QPO behavior at the initiation of the hard dip can ultimately be used to determine the terminating x-ray behavior, and the following infrared flaring behavior.

\end{abstract}


\section{Introduction}

Discovered in 1992 by \citet{cast92}, GRS~1915+105 is an x-ray transient that continues to intrigue us with its unique array of variability on many wavelengths and timescales. Dubbed a microquasar because of its apparent superluminal jets \citep{mirabel94}, GRS~1915+105 is a black hole candidate and x-ray binary. Although it suffers $\sim 20 - 30$ magnitudes of extinction at visible wavelengths, GRS~1915+105 has shown great activity in the radio, infrared, and x-ray regimes. Of particular interest in this study are x-ray light-curves showing spectrally-hard dips on $\sim 30$ minute timescales. 

It is well-established that the spectrally-hard dips that frequently appear in the x-ray light-curves of GRS~1915+105 are associated with infrared and radio flares \citep{eiken98, mirabel98, fender98, kleinwolt02}. These spectrally-hard dips are also associated with $2-10$~Hz variable quasi-periodic oscillations (QPOs). \citet{bell00} defines a total of twelve x-ray light-curve classes for GRS~1915+105, distinguishable by general appearance, count rate, and x-ray color (related to hardness).
We study the $\alpha$- and $\beta$-class light-curves in the $2 - 25$ keV range. Both have extended spectrally-hard dips but the length of the $\alpha$-class hard dip is nearly twice the length of the $\beta$-class. The $\alpha$-class light-curves have a dip of $\sim 1200$ seconds followed by strong x-ray oscillations marking the end of the hard state. The $\beta$-class light-curves are more complex. The dip lasts $500 - 700$ seconds and its ending is marked by a spectrally soft spike followed by a nearly monotonic rise. After reaching a peak flux, the x-rays begin rapid, large-amplitude oscillations. In both the $\alpha$- and $\beta$-classes, simultaneous x-ray and infrared observations show infrared flares rising as the hard x-ray dip ends \citep{eiken98, mirabel98, droth}. In the $\beta$-class, the infrared flare begins near in time to the soft spike, called the ``trigger spike''. \citet{eiken98} observed a one-to-one correspondence of x-ray dips to infrared flares at 30 minute intervals. 

While the x-rays are rapidly oscillating, the infrared flare peaks and decays without the rapid, large amplitude variations seen in the x-ray. Flares associated with our observations are defined by \citet{eiken00} as class~C and class~B, ranging from $30 - 200$ mJy when de-reddened by 3.3 magnitudes. Individual Class~C infrared flares are smaller ($5 - 10$ mJy) and have been observed individually associated with isolated x-ray flares in a dip/flare cycle \citep{eiken00}. When several class~C infrared flares occur in rapid succession, they can appear as a larger flare. \citet{droth} shows this to be the case for the $\alpha$-class light-curves in our sample, where the associated infrared flares range from $10 - 30$ mJy. 
\citet{droth} showed that if each soft x-ray flare were associated with a $5 - 10$ mJy (class~C) sub-flare then the duration and strength of the overall infrared flare would be explained. By that analysis, the predicted collection of infrared sub-flares associated with a $\beta$-class light-curve would contribute only a small fraction of the observed $60 - 200$ mJy infrared flux. This small contribution was observed by \citet{eiken00} as an ``infrared excess''. Although the period of the rise and decay of the primary class~B infrared flare is not coupled to the period of the x-ray oscillations, the duration of the infrared excess is \citep{eiken00,droth}. The difference between the dip/flare cycles associated with $\alpha$- vs. $\beta$-class curves becomes the presence of a large primary infrared flare which is uniquely associated with $\beta$-class x-ray light-curves. The presence of the primary flare is widely associated with the presence of the x-ray ``trigger spike''. It is believed that both are linked to the underlying cause of larger plasma ejections. 

Several authors have shown that infrared flares tend to be followed by radio flares of similar morphology \citep{fender98, mirabel98}. The sequence of a spectrally-hard dip, an infrared flare, then a radio flare, is generally associated with a plasma ejection from the source. The ejection is observed as the x-rays transition from a spectrally-hard to a spectrally-soft state. During the hard dip that precedes an ejection event, a $2 - 10$~Hz QPO is always observed \citep{bell00}. While the QPO is present, the disk x-ray emission is greatly reduced in the $2 - 25$ keV range and the x-ray luminosity is dominated by power law flux. Several groups have found the QPO peak frequency positively correlated to both power law and thermal disk components, suggesting that the QPO arises in the same location as the related emission or is causally related to it \citep{muno99,feroci99}. \citet{markwardt99} observed that at frequencies above 4~Hz, the QPO is most strongly correlated to the thermal disk component, specifically the blackbody disk flux. At lower frequencies, the QPO shows a broad association with the power law flux. \citet{muno99} also show the QPO is strongly associated with disk temperature during the hard dip. After the trigger spike, the power density spectrum becomes smooth \citep{markwardt99}.

In this paper, we examine the behavior of the QPO during the hard dip and the relationship of its behavior to subsequent infrared flares. In Section 2, we discuss our observations and analysis technique. We calculate and discuss the correlation of the QPO peak frequency to spectral features in Section 2.1 and comment on the time evolution of the QPO during the dip in Section 2.2. In Section 2.3, we examine the morphology of the trigger spike in the $\beta$-class light-curve and the indication of a continuum of behaviors from this class to $\alpha$-class, which does not show a trigger spike. In Section 2.4, we compare infrared flaring behavior between the events. We discuss the significance of these results in Section 3 and in Section 4 we summarize our conclusions.

\section{Observations and Analysis}

We use several {\it Rossi X-ray Timing Explorer} (RXTE) observations of GRS~1915+105 taken on 14 - 15 August 1997, 9 September 1997, and 27 - 28 July 2002 (see Table \ref{tbl-1}). We extract Proportional Counter Array (PCA) Standard-1 light-curves using FTOOLS v5.3 and identify regions showing a hard x-ray dip. We examine 12 regions in this work; nine of them are $\beta$-class and three are $\alpha$-class \citep{bell00}. They are numbered sequentially $\beta$-1 through $\beta$-9 and $\alpha$-10 through $\alpha$-12.
For the twelve regions, we extract binned mode 8-millisecond light-curves in the $2 - 13$ keV range and 4-second resolution binned and event x-ray spectra in the $2 - 25$ keV range.  Using XSPEC v11.3.0, we fit the spectra with a combination of absorbed multi-temperature disk blackbody and power law models. Only fits with $\chi_\nu ^2 < 2$ are used in analyses.

\begin{figure*}
%\epsscale{.80}
\plotone{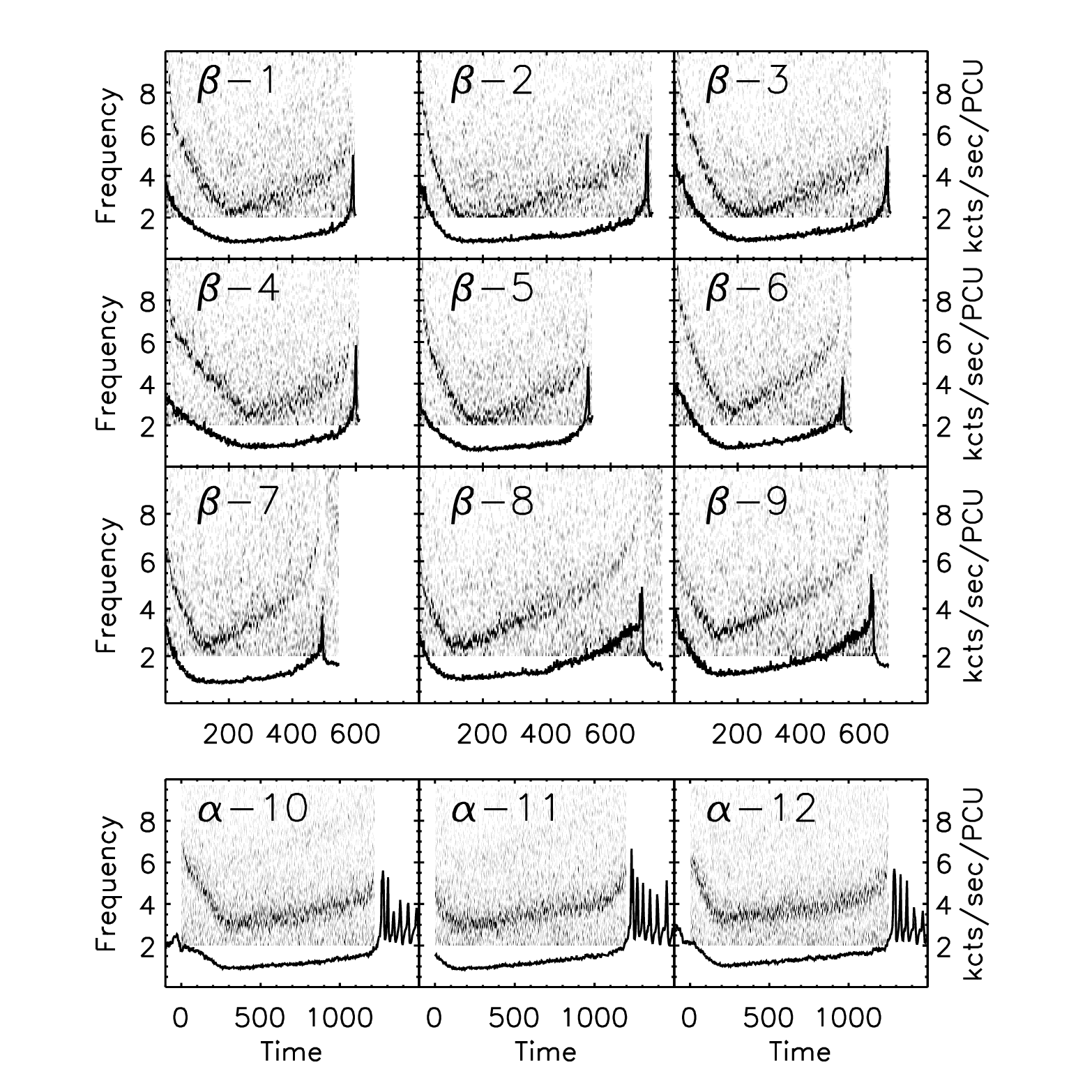}
\caption{A gray-scaled, fine-binned power density spectrum for the light-curves. We plot the 1-second resolution light-curve over the PDS. Note that, in the $\beta$-8 and $\beta$-9 cases, a strong low-frequency noise component appears $\sim$100 seconds before the spike.}
\label{fig1}
\end{figure*}

Using a Fast Fourier Transform, we calculate the power density spectrum (PDS) from the binned mode 8-millisecond light-curve. We fine-bin the PDS using Fourier interpolation and track the peak frequency at 4-second resolution \citep[see e.g.][]{ransom02}. Fine binning allows us to calculate a higher-resolution Fourier response by interpolating responses at non-integer frequencies. In Figure \ref{fig1}, we show a gray-scaled fine-binned PDS and overplot a one-second resolution light-curve.

\begin{figure*}
\plotone{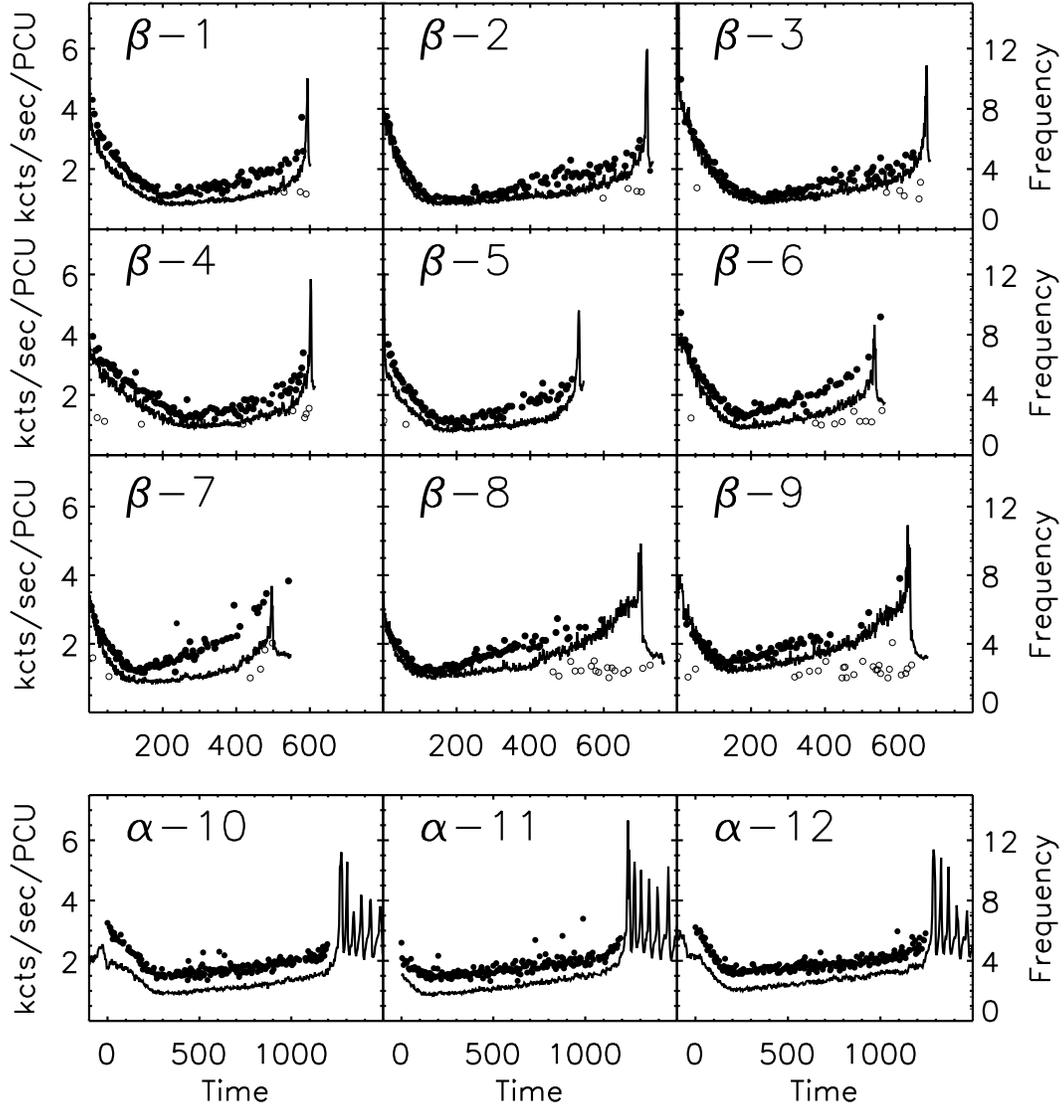}
\caption{An overlay of the 1-second resolution x-ray light-curve (line) and fine-binned 4-second QPO~frequency (circles). Time is in seconds and QPO~frequency is in~Hz. The open circles are low frequency power peaks observed when the QPO is not detected (see Section 2). They are shown for illustrative purposes. Note that for $\beta$-6 through $\beta$-9, the QPO spends less time near the minimum frequency.}
\label{fig2}
\end{figure*}

We determine the QPO~frequency by fitting a Moffat function to the PDS in the $2 - 10$~Hz frequency range. The Moffat function is a Lorentzian modified with a variable power law index \citep{moffat69}. 
The QPO is considered detected if it has a quality factor $Q=\nu / FWHM >2$, where $\nu$ is the centroid frequency of the Lorentzian and $FWHM$ is the full-width at half max. In $\alpha$-class curves, we require the QPO~frequency to be between 2.6 and 10~Hz to avoid contamination by low frequency noise. In $\beta$-class curves, we allow detections in the full $2 - 10$~Hz range. However, also visible in the PDS is a low-frequency noise component, the maximum frequency of which tends to rise above 2~Hz at the beginning and end of the dip. To avoid spurious detections from this component, we require that each detected QPO does not vary too sharply from the QPOs at surrounding times. For points at times $t > 350$~seconds, we take an average of the prior 10 detected frequencies, $\nu _{avg}$. The next detection is required to be greater than 75\% of $\nu _{avg}$. For $t < 150$ seconds, QPOs are treated similarly, though $\nu_{avg}$ is determined by using points following the reference point, as opposed to those preceeding it. Although omitting points as non-detections can leave part of the QPO evolution under-sampled, we believe we have sufficient representation from the regions to ensure a sound qualitative result and a reasonable quantitative one. In addition this method yields a consistent and repeatable QPO detection. It is possible that adjusting the range in this manner causes us to omit real, unresolved oscillations in the QPO, but visual inspection shows that these points are not part of the primary U-shaped QPO feature we wish to focus on. This shape is apparent in the PDS shown in Figure \ref{fig1}. In Figure \ref{fig2}, we plot the detected QPOs  as filled circles against a 1-second resolution light-curve. For illustrative purposes, we show power peaks in the PDS that do not meet the detection criteria as open circles. 

\subsection{QPO~Frequency - Correlation to Spectral Features}

 For each hard dip, we calculate the Linear Pearson Correlation Coefficient, $r$, between the QPO~frequency and various spectral features. We define events with $|r|>0.70$ as highly correlated. This value is chosen as it means that at least half the variance of the spectral feature can be accounted for by the variance of the QPO~frequency. Values of $|r|$ between 0.4 and 0.7, we discuss as believable, but generally disregard when more statistically significant correlations are present. 

We list the correlation coefficients for the twelve dips in Table \ref{tbl-2}. In all of the cases, we observe a highly significant correlation between the QPO~frequency and the total flux which is generally stronger than those to individual blackbody or power law features. When considering model-specific spectral features, eleven out of twelve show strong correlations to power law flux, while only six out of those eleven show strong correlations to blackbody features. In nine out of those eleven cases, the correlation to the power law flux is stronger than the correlation to any blackbody feature. This gives the apparent result that the QPO~frequency is more fundamentally tied to the power law component than the blackbody component. 

Figure \ref{fig3} shows a scatter plot of the QPO~frequency versus the power law flux for the twelve cases. The points marked with triangles are detections within the first 100 seconds of the dip and the squares are from the last 100 seconds of the dip. The line is based on a fourth order fit to the time evolution of the QPO~frequency and the power law flux, tracing the approximate path of the evolution. The open circles are power peaks between 2 and 10~Hz where a QPO was not detected (see above). These points are generally not a part of the observed trends. 

In $\beta$-1 through $\beta$-5 (August 1997) the evolution of the QPO-power law flux relation is generally tighter and the correlations stronger ($r \ge 0.80$). The $\alpha$-10 through $\alpha$-12 curves also show a tight correlation, though the QPO frequency varies over a smaller range of frequencies. Like the $\alpha$-class curves, we only have detections for $\beta$-8 over a small range of frequencies, during which we see a tight correlation to the power law flux. In the other three cases, $\beta$-6, $\beta$-7, and $\beta$-9, we notice a hysteresis effect, observed as a separation between the entrance (triangles) and exit (squares) points of the dip. In the $\beta$-7 case, the hysteresis is extreme and lowers the correlation coefficient drastically despite the initially linear correlation at the dip entrance. The squares which trace the last 100 seconds of the dip show the QPO~frequency rising sharply while the power law flux is relatively constant, suggesting a decoupling of the features.

In $\beta$-1 through $\beta$-7, the correlation between the QPO~frequency and the power law flux is accompanied by a similarly strong correlation to the blackbody flux and blackbody disk temperature. In Figure \ref{fig4}, we see a strong correlation to the blackbody flux is most prominent above 4~Hz, a point also observed by \citet{markwardt99}. However, the correlation does not last over the broad part of the hard dip where the QPO can change significantly while at relatively constant blackbody flux. In $\beta$-8 and $\beta$-9, where the blackbody flux appears uncorrelated, there are fewer QPO detections above 4~Hz due to the rise in low frequency noise at the dip exit. 

We show the QPO~frequency - blackbody temperature scatter plots in Figure \ref{fig5}. Clearly the QPO in the $\alpha$-class curves is not believably correlated to the blackbody temperature. In $\beta$-1 through $\beta$-5 and $\beta$-8, the correlations to the blackbody temperature follow a steady trend, but suffer a slightly wider dispersion in their evolution than the power law flux. In $\beta$-6, $\beta$-7, and $\beta$-9, we see a similar hysteresis to that observed in the power law flux. To untangle the possible interplay of the blackbody temperature and power law flux, we apply a partial correlation analysis. The partial correlation coefficients are listed in Table \ref{tbl-3}. In the table, we calculate coefficients for four scenarios:
\begin{enumerate}
\item The QPO~frequency --- power law flux correlation removing the effect of blackbody flux.
\item The QPO~frequency --- power law flux correlation removing the effect of blackbody temperature.
\item The QPO~frequency --- blackbody flux correlation removing the effect of power law flux.
\item The QPO~frequency --- blackbody temperature correlation removing the effect of power law flux.
\end{enumerate}

In the first scenario, we see that with the blackbody flux removed, the correlation to power law flux is still strong. This is expected because the blackbody flux poorly explains variances at frequencies below 4~Hz (see Fig. \ref{fig4} and Table \ref{tbl-2}). In the second case, we find the removal of blackbody temperature has a more significant effect. In most cases, the QPO~frequency --- power law flux correlation is weak, though believable ($r > 0.4$). This means that after removing variations in the QPO~frequency and power law flux that can be explained by variations in blackbody temperature, variations in the QPO~frequency remain that can be at least partially explained by variations in the power law flux. High significance is seen in $\alpha$-class cases, which is expected due to their weaker dependence on blackbody temperature. In the third case, we test the correlation to the blackbody flux after removing the power law flux. These correlations are believable and occasionally strong, suggesting that a combination of the blackbody and power law sources is required to explain the QPO~frequency. In the final case, the removal of power law flux from the QPO~frequency --- blackbody temperature correlation, we see that partial correlation coefficient drops below significance in half of the cases. This means that once variations in the QPO~frequency that can be explained by variations in the power law flux are removed, no variation remains that can be explained by blackbody temperature. 

 In summary, the QPO~frequency is  often correlated to the power law flux. For $\beta$-class light curves, when this correlation is strongest, we tend to find a correlation to the blackbody flux and blackbody temperature as well. In cases where there is a slightly weaker correlation to the power law flux, the correlation to the blackbody features is less predictable and a hysteresis effect is visible in the power law flux and blackbody temperature relations. In contrast, $\alpha$-class QPOs tend to have a strong correlation to the power law flux and weak or non-existent correlations to the blackbody features. Based on this evidence, we believe the correlation between the QPO~frequency and the power law flux is more fundamental.

\subsection{QPO Time Behavior}

\begin{figure*}
\epsscale{.8}
\plotone{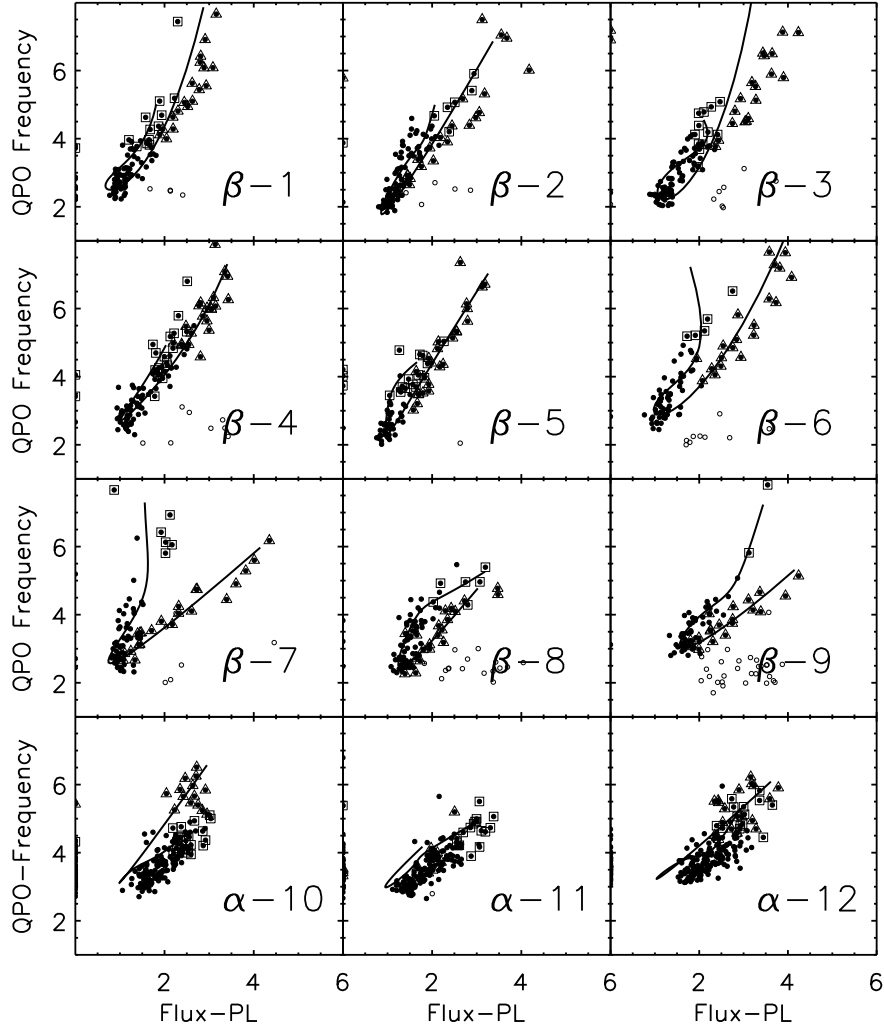}
\caption{Scatter plots of QPO~frequency (in~Hz) with power law flux, Flux-PL, (in $10^{-8} erg~cm^{-2} s^{-1}$)  for $\alpha$- and $\beta$-class light-curves. Triangles indicate detections in the first 100 seconds of entering the dip. Squares are points within the last 100 seconds before exiting the dip. The line is a fourth-order best fit to the time evolution of the features.  Note that for $\beta$-1 through $\beta$-5, a strong linear correlation is apparent. For $\beta$-6, $\beta$-7, and $\beta$-9, the correlation weakens and we see different degrees of hysteresis.  The open circles (non-detections of the QPO as defined in Figure \ref{fig2}) do not contribute to the apparent hysteresis pattern. In $\beta$-8, the range of frequencies is significantly less, probably due to increased non-detections. This is evidence that a continuum of QPO behaviors exists within the $\beta$-class light-curves. In $\alpha$-class light curves, a correlation exists over a smaller range of frequencies as well.}
\label{fig3}
\end{figure*}

\begin{figure*}
\epsscale{.8}
\plotone{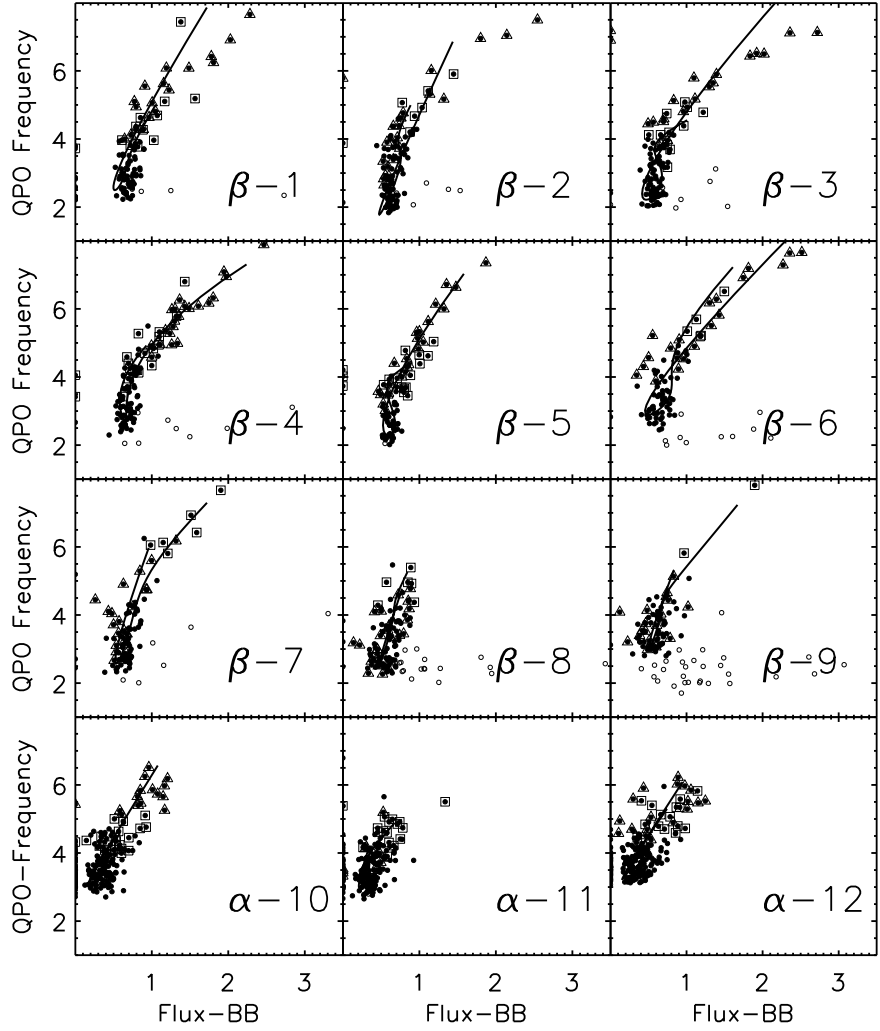}
\caption{Scatter plots of QPO~frequency (in~Hz) with blackbody flux, Flux-BB, (in $10^{-8} erg~cm^{-2} s^{-1}$) for $\alpha$- and $\beta$-class light-curves. Symbols are as in Figure \ref{fig3}. A correlation is observed above 4~Hz, but below 4~Hz, the QPO can change significantly while the blackbody flux remains relatively constant.}
\label{fig4}
\end{figure*}

\begin{figure*}
\epsscale{.8}
\plotone{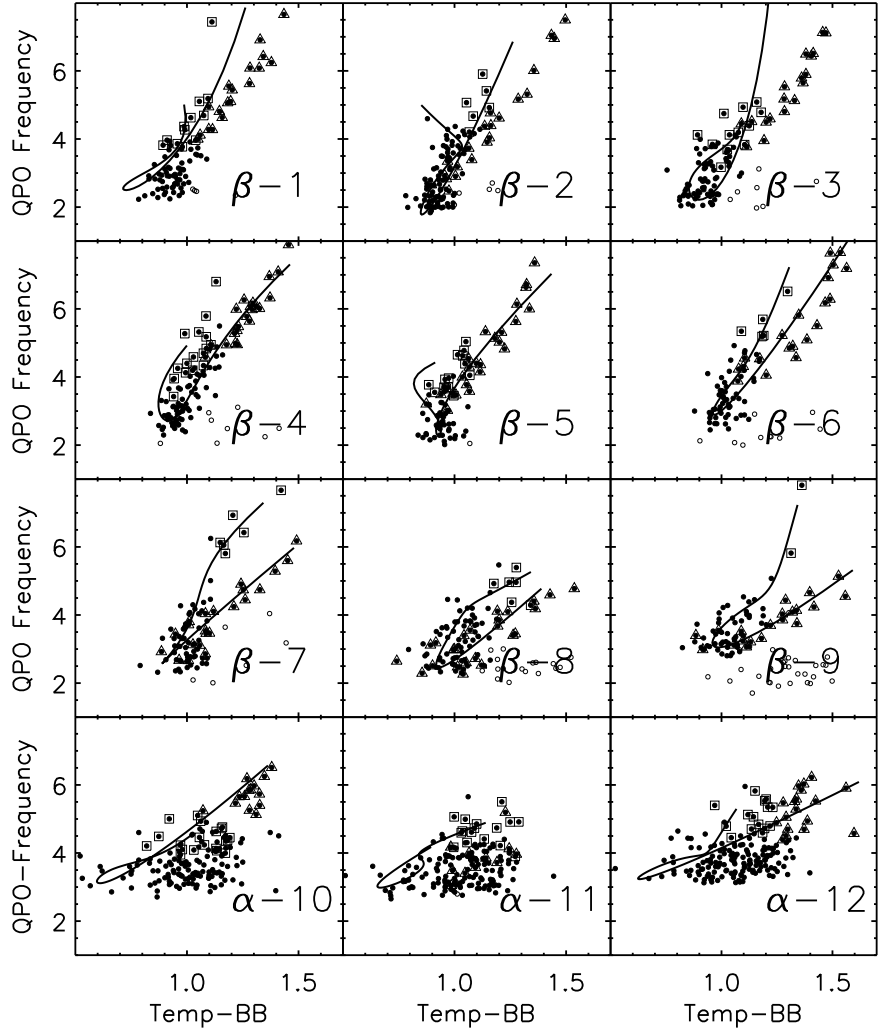}
\caption{Scatter plots of QPO~frequency (in~Hz) with blackbody temperature, Temp-BB, (in keV) for $\alpha$- and $\beta$-class light-curves. Symbols are as in Figure \ref{fig3}. The $\alpha$-class light curves clearly show no correlation. In most of the $\beta$-class light curves, a correlation is seen but with a wider dispersion than that of the power law flux relation. Hysteresis is still apparent in $\beta$-7 and $\beta$-9 particularly.}
\label{fig5}
\end{figure*}

\begin{figure*}
\epsscale{.80}
\plotone{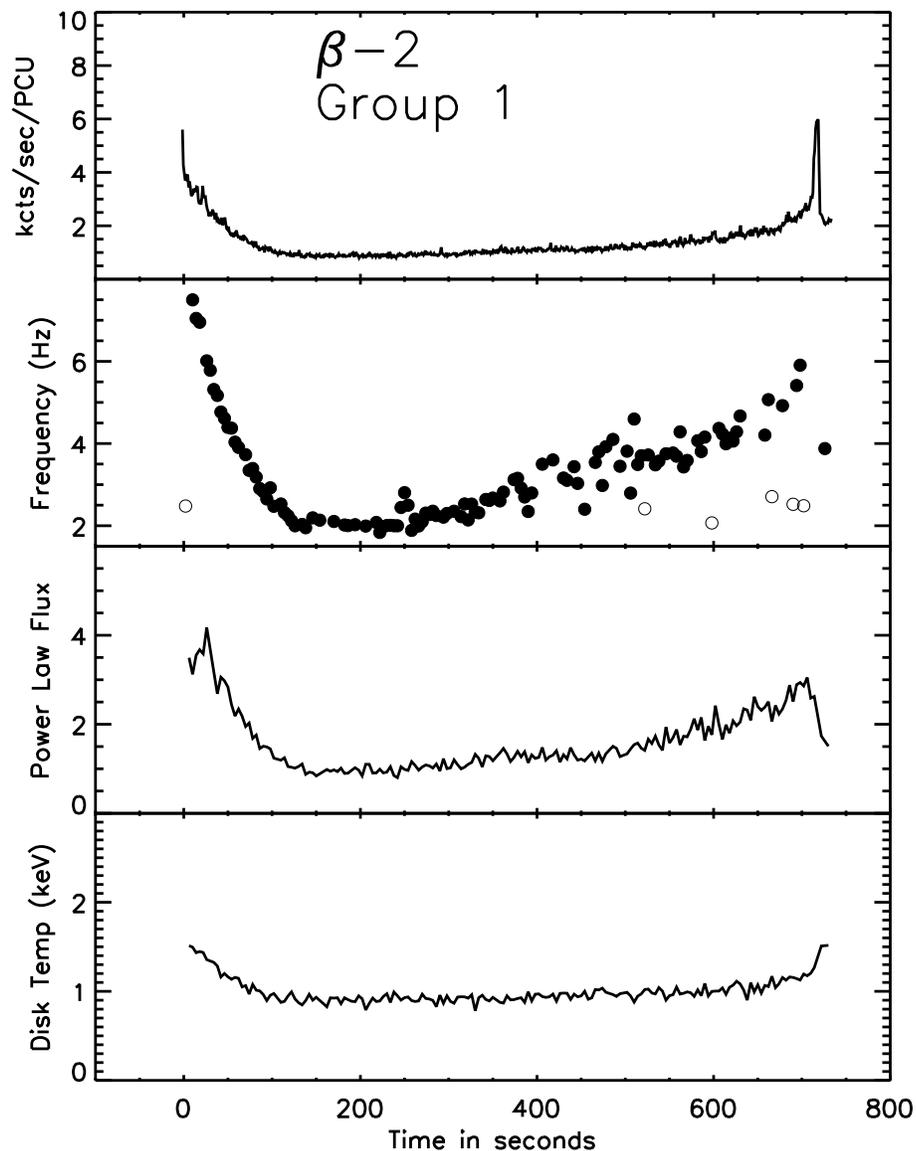}
\caption{The time evolution of a Group~1 event. The top panel shows a 1-second resolution x-ray light-curve. The next panel shows the fine-binned QPO~frequency and the open circles are low frequency power peaks where the QPO is not detected. The x-ray power law flux (in units of $10^{-8} erg~cm^{-2} s^{-1}$) and blackbody disk temperature are shown at four-second resolution. The QPO~frequency is most strongly correlated to the power law flux during the hard dip, but is also correlated to the blackbody disk temperature.}
\label{fig6}
\end{figure*}

\begin{figure*}
\epsscale{.80}
\plotone{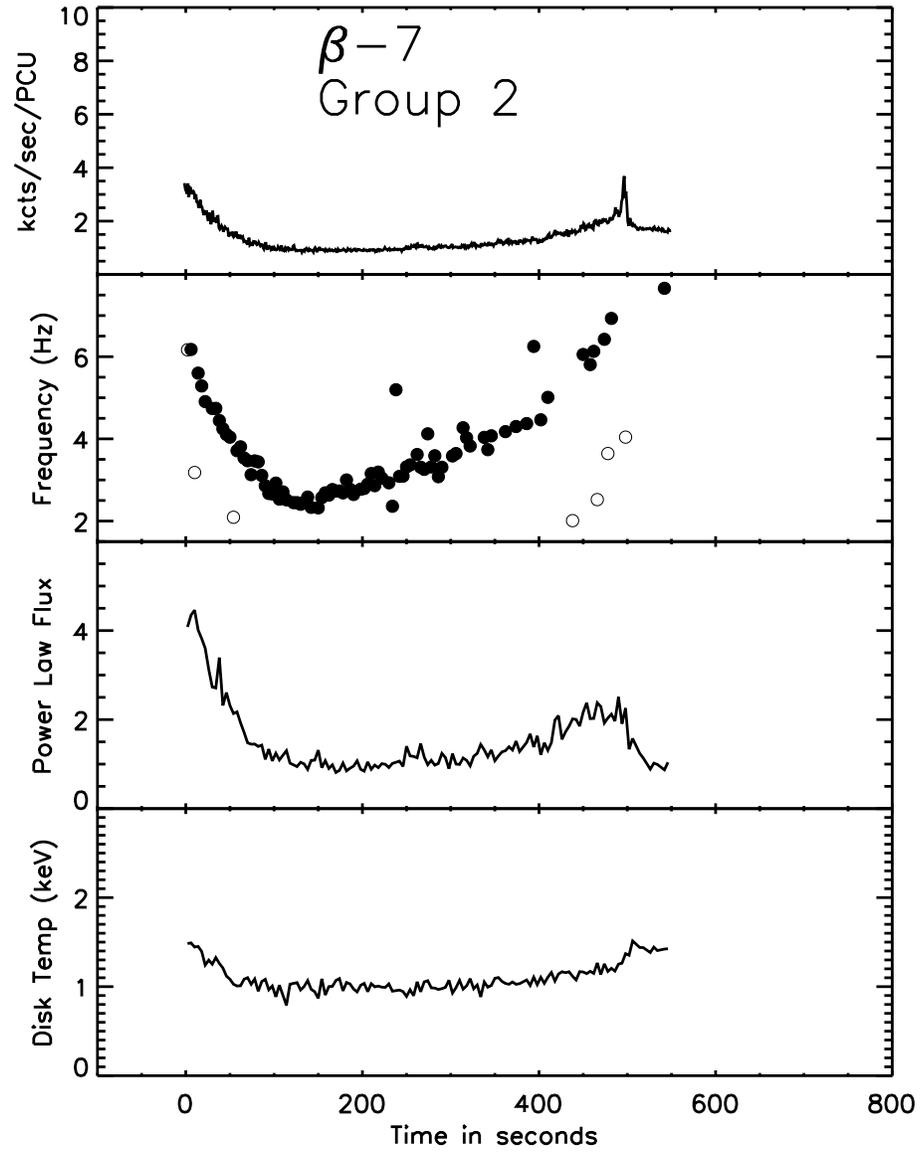}
\caption{The time evolution of a Group~2 event. Panels are as described in Figure \ref{fig6}. In this case, the QPO~frequency begins rising while the power law flux and the blackbody temperature are relatively constant.}
\label{fig7}
\end{figure*}

\begin{figure*}
\epsscale{.80}
\plotone{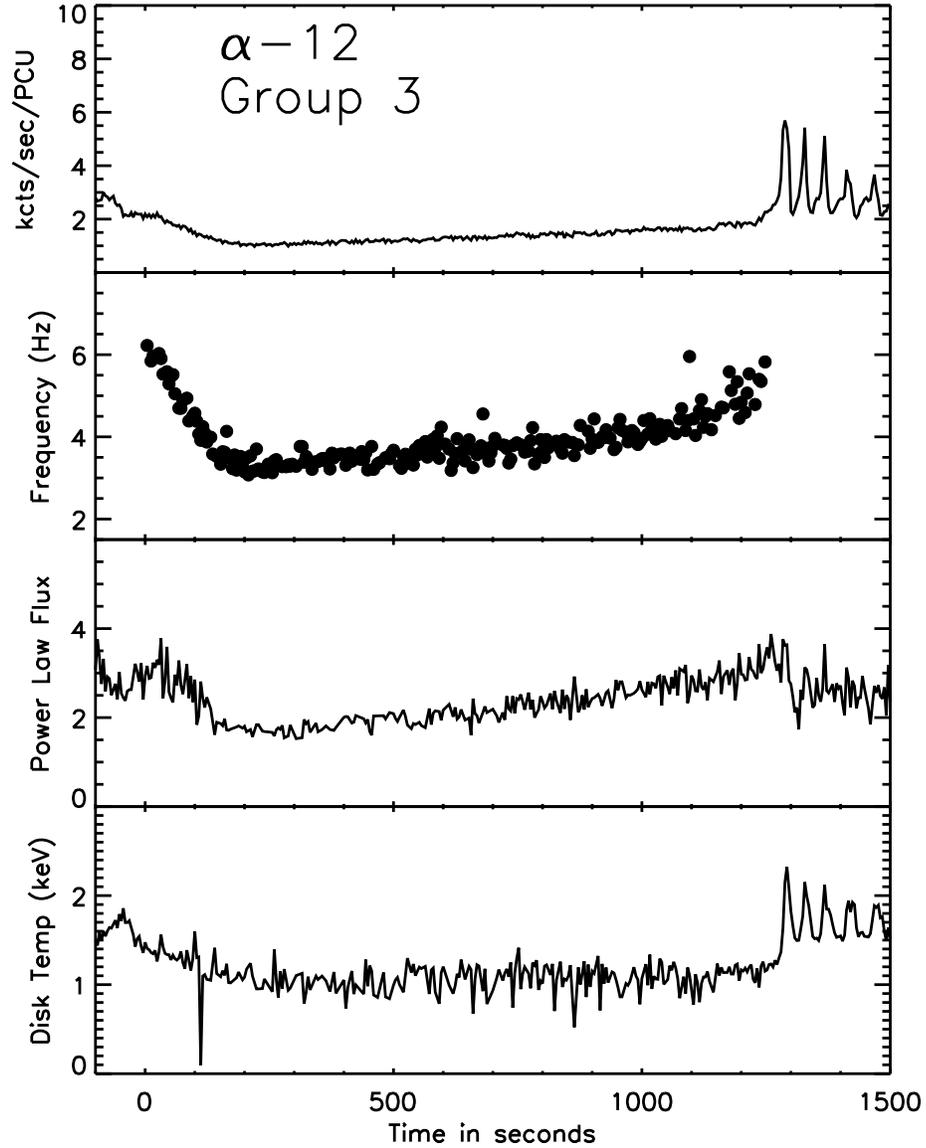}
\caption{The time evolution of a Group~3 event. Panels are as described in Figure \ref{fig6}. The hard dip lasts $\sim 1200$ seconds ---  twice the length of the Group~1 and 2 events. The QPO~frequency behavior is similar in morphology to Group~1, and while strongly correlated to the power law features, there is no accompanying correlation to blackbody features.}
\label{fig8}
\end{figure*}

Noting the hysteresis in the QPO~frequency versus power law flux distribution in several of the $\beta$-class light curves (see Fig. \ref{fig3}), it is likely that the spectral evolution is intimately tied to the QPO time evolution. The QPO evolution at the beginning of all the hard dips is similar, each beginning with the sudden appearance of a QPO at $6 - 10$~Hz. This QPO will smoothly drop to between $2 - 3$~Hz within seconds of the initiation of the hard dip. After the x-ray trigger spike, the primary U-shaped QPO feature will disappear, replaced by occasional power peaks. In Table \ref{tbl-4} we list the minimum QPO~frequency during the dip and the time spent near that frequency. Based on the observable variations, we then divide the light-curves into three broad groups:
\begin{itemize}
\item {\bf Group~1}: Figure \ref{fig6} shows an example of the $\beta$-class light-curves in Group~1. A series of x-ray oscillations calm into a low, spectrally-hard dip within about 100 seconds. During this time, a QPO arises at $\sim 6-8$~Hz. Following the intensity drop of the light-curve and more specifically the power law flux, the QPO falls steadily to $\sim 2$~Hz. It remains at this frequency for over 150 seconds (see Table \ref{tbl-4}), after which the power law flux and QPO~frequency begin a slow rise. The U-shaped QPO vanishes after the x-ray trigger spike. The total length of the dip is on the order of 600 seconds, and $\sim$ 30\% of that time is spent at the minimum frequency.
\item  {\bf Group~2}: This group is also composed of $\beta$-class light-curves with similar spectral behavior to Group~1 (see Fig. \ref{fig7}). However, the QPO behavior in this group is somewhat different. While in Group~1, the QPO falls off to $\sim 2$~Hz and lingers, in Group~2 the QPO immediately starts to rise again. In Table \ref{tbl-4}, we show that while the Group~1 events remain near the minimum frequency for $> 150$ seconds, the Group~2 events remain for $< 100$ seconds (about 15\% of the dip length). The following rise in frequency is the start of the hysteresis in the QPO~frequency --- power law flux scatter plot (Fig. \ref{fig3}) and likely indicates that the QPO and power law flux have decoupled. In addition, these cases see the rise of a low frequency noise component above 2~Hz as the QPO weakens in amplitude and rises rapidly in frequency. While these low frequency points are excluded from correlation analysis as being associated with low frequency noise, it is possible that they represent an increase in rapid, unresolved QPO oscillations. The difference in behaviors in the first two groups is most remarkable because their x-ray light-curves and spectral behaviors are so similar.

\item  {\bf Group~3}: The final group contains the $\alpha$-class light-curves represented in Figure \ref{fig8}. In these events, the hard dip is surrounded by x-ray oscillations but no independent terminal spike is observed. The total length of the dip is $\sim 1200$ seconds and the QPO disappears when the dip ends. This case is similar in shape to Group~1 QPO evolution, though twice as long. It differs in that on entering the dip the QPO~frequency levels off at $\sim 3$~Hz. Overall, the frequency varies over a smaller range than seen in the other two groups.  
\end{itemize}

\subsection{Differing Trigger Spike Morphology in $\beta$-class Light-Curves}

\begin{figure*}
\plotone{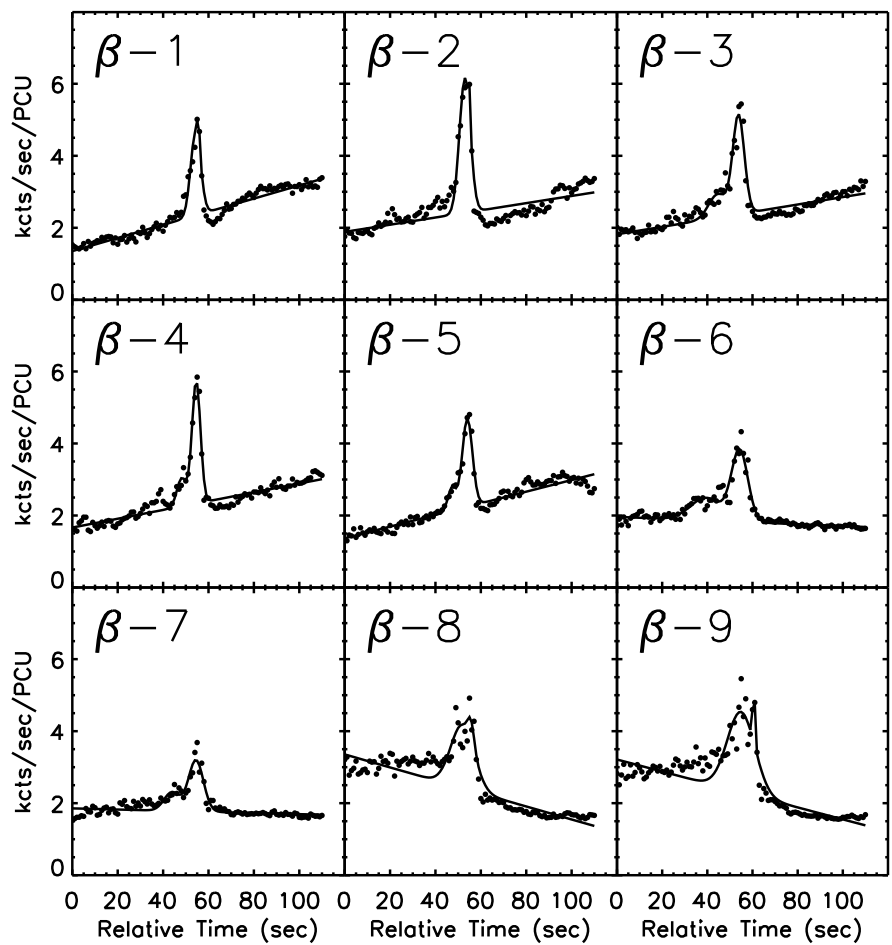}
\caption{The trigger spike of the $\beta$-class light-curves for 1997 August ($\beta$-1 through $\beta$-5) and 1997 September ($\beta$-6 through $\beta$-9) data at one-second time resolution. The solid lines are a double Gaussian plus polynomial fit to the data points. We classify the $\beta$-1 through $\beta$-5 events as Group~1. These have a strong, narrow, symmetric spike and the underlying flux has a positive slope. The $\beta$-6 through $\beta$-9 data we classify as Group~2 events. These have weaker, wider, asymmetric spikes and the underlying slope is flat or negative. The $\beta$-8 and $\beta$-9 light curves have more disturbed spike morphologies on this time scale.}
\label{fig9}
\end{figure*}

The separation of the $\beta$-class curves into two groups of QPO evolution inspires the search for other differences in the light-curve behavior. We focus here on the trigger spike which appears at the end of the hard dip and coincides with the start of the infrared flare and note that the intensity and morphology of the spikes vary between Group~1 and Group~2 events. Figure \ref{fig9} shows a one-second resolution light-curve for each of the nine observed spikes in a time range of 55 seconds before and after the maximum.

The light-curves are fit with a double Gaussian plus a polynomial of the form:
\begin{displaymath}
f=A_0 e^{-z_A^2} + B_0 e^{-z_B^2} + C_0 + C_1 t
\end{displaymath}
where $z_A=(t - \tau _A)/\sigma _A$ and $z_B= (t - \tau _B)/\sigma _B$. 

The second Gaussian term fits a small rise preceeding the x-ray spike --- a feature more distinct in the $\beta$-3 through $\beta$-7 light-curves. This second Gaussian is treated with the polynomial as part of the background emission. A normalized peak amplitude for the spike is calculated as $A= A_0/A_{bg}$ where $A_{bg}$ is the polynomial and second Gaussian evaluated at the primary Gaussian peak time, $t=\tau_A$. It should be noted that the apparent symmetry of the spike is tied to the time resolution. The one-second time resolution allows a fair sampling of data points for the range while smoothing the rapid x-ray variability seen in an eight-millisecond curve. Broader time intervals will leave the spike under-sampled and reduce the accuracy of the fit.

 The relevant numerical fits are listed in Table \ref{tbl-5}. It is interesting to note that the integrated count rate ($f_{int}$) over the full-width-half-max of the peak is similar for all data sets, but in most other aspects of the fit, the August (Group 1) and September (Group 2) data have different properties. 

\begin{itemize}
\item The Group~1 data have a higher normalized amplitude: $A^{Grp 1} > A^{Grp 2}$. 
\item The Group~1 data are more symmetric while the Group~2 data shows a sharp cut-off after the peak flux (see Fig. \ref{fig9}).
\item The Group~1 data spikes are narrower: $\sigma _A^{Grp 1} < \sigma _A^{Grp 2}$.
\item The underlying slope of the light-curve is positive in the Group~1 data and flat or negative in the Group~2 data: $C_1^{Grp 1} > 0$ while $C_1^{Grp 2} < 0$.
\end{itemize}

While all Group~2 events show a negative underlying light-curve slope, it is interesting that two of them ($\beta$-6 and $\beta$-7) have a nearly flat slope while the other two ($\beta$-8 and $\beta$-9) have a decidedly negative slope. From Figure \ref{fig3} we note that the two with nearly flat slopes have a slightly more visible hysteresis because they have more QPO detections above 4 Hz. The low frequency noise component does not rise as strongly above 2~Hz and they show a stronger correlation to blackbody temperature. In addition, the terminating spikes of $\beta$-8 and $\beta$-9 look much more disturbed than those of other light-curves (Fig. \ref{fig9}). The differences in trigger spike morphology suggest that the variation in QPO behavior is not an artifact of the detection method. The $\beta$-6 event is particularly interesting because it can be considered a crossover between Group~1 and Group~2 events.  The QPO~frequency --- power law flux relation in $\beta$-6, shows a significant correlation, $r=0.78$, despite the apparent hysteresis. From Figure \ref{fig3}, we see that this event seems to be a bridge between the sharp linear correlations of Group~1 and the divergent shapes of Group~2. We classify it as Group~2 because the QPO~frequency clearly deviates from the initial regression and because of its trigger spike morphology. The trigger spike of $\beta$-6 is wider and more asymmetric than Group~1 events. In addition, the flat underlying light-curve of $\beta$-6 suggests that it is more appropriately associated with Group~2 than Group~1 events. This being said, we acknowledge that the groups are not absolute, but likely part of a continuum of behaviors. 

\subsection{Associated Infrared Flaring}

Simultaneous infrared coverage is available for four out of five of our Group~1 data sets. \citet{eiken98} show that these dip/spike pairs are usually followed by large infrared flares. The events observed ranged from $\sim 60$ to 200 mJy. Although the rise and fall of the flare does not correspond to the period of x-ray oscillation, a weak infrared excess which lasts throughout the period of the x-ray oscillations is observed. \citet{eiken00} and \citet{droth} explain this excess as the superposition of many faint infrared flares each on the order of $\sim 10$ mJy. The dominant infrared flare is associated with the x-ray trigger spike.

\citet{mirabel98} observed a single infrared flare event associated with our $\beta$-7 curve which reached an amplitude of $\sim 30$ mJy. Simultaneous infrared coverage is not available for the other hard dips in Group~2. In all three $\alpha$-class light-curves of Group 3, the hard dip is followed by a $\sim 30$ mJy infrared flare \citep{droth}. In \citet{mirabel98}, an $\alpha$-class event is observed to be followed by a flare that peaked at $\sim 10$ mJy. \citet{droth} showed that the $\sim 30$ mJy flares can be explained as a summation of Class~C sub-flares, each associated with a soft x-ray flare.

\section{Discussion}
\subsection{QPO Correlation with Spectral Features}

Previous studies have shown that the QPO is most strongly tied to the thermal disk component \citep{markwardt99, feroci99, muno99}. Using the September~1997 data (our $\beta$-6 through $\beta$-9), \citet{markwardt99} point out that the correlation to disk flux is strongest when the QPO~frequency is above 4~Hz. While this is true, the QPO~frequency is in this range less than 25\% of the time, and mostly falls into this range when entering or exiting the hard dip. During the course of the hard dip, the QPO will change significantly while the blackbody flux remains relatively constant. \citet{markwardt99} also say that at lower frequencies, there is an apparent broad correlation with power law flux. We confirm that this correlation may exist, and show that it is most apparent at lower frequencies. Because of a hysteresis effect, the deviation in the QPO frequency --- power law flux relationship is more apparent at high frequencies. We suggest that an initially tight correlation is broken as the QPO begins to rise in the latter half of the dip. Two out of four of the observed September events are strongly correlated to power law flux and a different two out of four are correlated to blackbody temperature.

 In contrast, the August 1997 (our $\beta$-1 through $\beta$-5) events show a strong correlation with both power law and blackbody features - specifically the power law flux and blackbody temperature. A partial correlation analysis shows that if either the power law flux or blackbody temperature is removed, the correlation to the other is weakened, so it is not likely the effects of these components can be untangled. We do, however, argue that the correlation to the power law component may be more fundamental, especially since the power law flux is more strongly tied to the x-ray emission at this point and the disk component is vanishingly small. In addition, the $\alpha$-class curves show a consistently strong correlation to the power law flux and less consistent correlation to blackbody features. 

As mentioned before, the bulk of the QPO change occurs during the entry into and exit from the dip. While in the dip, the spectral features remain fairly stable. Thus much of the correlation strength is tied to QPO stability during the dip and the relatively short motion at the start and end of the dip. This being the case, the rise in low-frequency noise which greatly affects QPO detections in $\beta$-8 and $\beta$-9 is also likely to be affecting the correlation coefficients. It is reasonable to assume that the number of non-detections late in the dip is reducing the number of QPO detections above 4 Hz and thus artificially lowering the correlation to blackbody features in these two cases. Short time-scale variations (where the QPO suddenly dips and recovers) also tend to create outliers to the correlation because the change in frequency is not accompanied by spectral variation. It is uncertain if these jumps are real as the shape of the variation is indeterminant on four-second time scales. These outliers are omitted from our correlation analysis.

\subsection{Infrared Flaring Behavior}
 When comparing these events, we cannot ignore the fact that all these hard dips are followed by an infrared flare, the strength of which is tied to the presence or absence of a terminating spike. Furthermore, the shape of the spike is tied to the behavior of the QPO during the preceding dip. Therefore, the properties of the QPO appear to be fundamental for determining the subsequent jet activity in GRS 1915+105, and the question becomes, ``If all hard dips start relatively the same (with the appearance of a QPO), why don't all hard dips end the same?'' What causes the QPO to quickly rise in Group~2 or to have a higher minimum frequency in Group~3? Is the energy that would be fed into the QPO and subsequent infrared flare for a Group~1 event being leaked out before the dip termination in Group~2 events, or is the energy missing from the system altogether? The Group~1 events have strong, symmetric trigger spikes and strong $\sim 100$ mJy infrared flares. The relatively weak trigger spike and $\sim 30$ mJy flare observed in September 1997 may be evidence that a weaker trigger spike would be associated with a weaker primary flare. 

On a final note, what we have referred to as the ``trigger spike'' of the $\beta$-class is so named because it coincides with the start of the infrared flare. However, observations of several $\beta$-class events by \citet{eiken98} were unable to conclusively determine whether the infrared flare began simultaneously with the trigger spike. Also, in looking at the \citet{mirabel98} event corresponding to our $\beta$-7, one might believe that the infrared flare starts $100 - 200$ seconds PRIOR to the spike \citep[see][their Figure 3]{mirabel98}. Noting that the QPO also significantly weakens compared to the low frequency noise component $100 - 200$ seconds prior to the spike (see Fig. \ref{fig1}) suggests that the  {\it origin} of the infrared flare may be causally linked to the mechanism powering the QPO, and not tied initially to the soft x-ray flaring or trigger spike. This suggests that the QPO is tied to a multi-wavelength energy release or the formation of multi-wavelength features. The decoupling of the QPO from x-ray spectral features in this case supports this. In all other cases, the disappearance of the QPO coincided with the trigger spike or the first x-ray oscillation (and thus the infrared flare), so this picture would be consistent. \citet{eiken00} observed a series of class~C infrared flares preceding the x-ray soft flares in $\theta$-class dip-flare cycles, so this sequence would not be entirely unprecedented.

\subsection{A Cause And Effect Summary}

While the nature of the QPO is still uncertain, it does reflect and may possibly be used to predict observable outcomes. In all of these cases, a QPO appears when the x-ray energy changes from soft to hard. While the change in hardness occurs very quickly, the flux drops slowly for $\sim 100$ seconds. The QPO, initially around 6~Hz, falls with similar smoothness (see Fig. \ref{fig2}). This is where the tracks diverge. We believe that the QPO behavior at the divergent point can ultimately be used to predict how the dip will end. Consider the three groups we identify, summarized in terms of cause and effect:

\begin{itemize}
\item {\bf Group~1}: A $\beta$-class light-curve enters a hard dip phase. The QPO falls off to 2~Hz and maintains that frequency for $> 150$ seconds.
\newline{\bf END RESULT}:
      \begin{itemize}
	\item The QPO~frequency is tightly correlated to both blackbody and power law spectral features and the correlation lasts the length of the dip.
	\item The dip lasts $500 - 700$ seconds and terminates with a strong, narrow, symmetric spike and increasing underlying flux.
	  \item A strong class~B infrared flare of strength $\sim 100$ mJy follows with infrared excess explained by pursuant x-ray oscillations.
      \end{itemize}
\item {\bf Group~2}: A $\beta$-class light-curve enters a hard dip phase. The QPO falls off to 2~Hz, but begins increasing after $< 100$ seconds.
\newline{\bf END RESULT}:
      \begin{itemize}
	\item The QPO~frequency decouples from the spectral features and the strength of the correlation is related to the degree of hysteresis.
	\item The QPO significantly weakens before the terminating spike.
	\item The dip lasts $500 - 700$ seconds and terminates with a weak, wide, asymmetric spike and flat or decreasing underlying flux.
	\item A slightly weaker class~B infrared flare of strength $\sim 30$ mJy follows, possibly starting with the weakening of the QPO.
     \end{itemize}
\item {\bf Group~3}: An $\alpha$-class light-curve enters a hard dip phase. The QPO falls off to 3~Hz and maintains that frequency for several hundred seconds.
\newline{\bf END RESULT}:
      \begin{itemize}
	\item The QPO is correlated to power law features but not to blackbody features.
	\item The dip lasts $\sim 1200$ seconds and then immediately enters a period of oscillation.
        \item A series of class~C infrared flares with summed amplitude $\sim 10 - 30$ mJy follows and is associated with the duration of the x-ray oscillations.
     \end{itemize}
\end{itemize}

\section{Conclusions}
In conclusion, we have identified three groups of hard dips with a range of spectral behavior, QPO~frequency evolution, trigger spike morphology, and infrared flare strength. Each hard dip is associated with a variable low frequency QPO and is followed by an infrared flare indicating an ejection event. Similarities in the x-ray light-curve, x-ray hardness, and infrared flaring suggest that a similar mechanism is responsible for these behaviors. 

While it is easy to expect different behavior from $\alpha$- and $\beta$-class light-curves, it is surprising to find differences within the $\beta$-class itself. Most interesting is the spectrum of QPO time evolution behaviors seen in our small set of observations and the fact that only a slight variation is necessary to affect the end result. The correlation of the QPO to x-ray spectral features hinges on the time evolution of the QPO. The time evolution is also intimately tied to the trigger spike morphology and subsequent infrared flaring. Most interesting is the possibility that the infrared flare begins with the disappearance of the QPO and is not solely tied to the trigger spike or x-ray flares. It is clear that this study only scratches the surface of the spectrum of behaviors exhibited by GRS~1915+105, however, by studying these events together, we may better understand the underlying mechanism. 

\acknowledgments 

The authors owe a debt of gratitude to the astute referee for aiding in the refinement of this work. V.~J.~M. and S.~S.~E. are supported in part by an NSF CAREER Award (AST-0328522). V.~J.~M. is also supported by a University of Florida Alumni Fellowship. D.~M.~R. is supported by a National Science Foundation Graduate Research Fellowship. This research has made use of data obtained from the High Energy Astrophysics Science Archive Research Center (HEASARC), provided by NASA's Goddard Space Flight Center.

\clearpage
%figures

% qpoalign

% qpo-plf scatter

% qpo-fbb scatter

% qpo-tbb scatter

% adding spectral time evolution

% gaussian fits

% tables
\begin{deluxetable}{cccc}
\tablecaption{{\bf XTE Observations}}
\tablewidth{0pt}
\tablehead{ \colhead{ID}  & \colhead{XTE DATA ID} & \colhead{Date Observed}& \colhead{Start Time (UTC) }}
\startdata
 $\beta$-1    & 20186-03-03-01 & 1997 Aug 14        & 04:20:52 \\
\tableline     
 $\beta$-2    & 20186-03-03-01 & 1997 Aug 14        & 05:50:52  \\
\tableline     
 $\beta$-3    & 20186-03-03-01 & 1997 Aug 14        & 07:17:12  \\
\tableline     
 $\beta$-4    & 20186-03-03-01 & 1997 Aug 14        & 09:18:56  \\
\tableline\tableline
 $\beta$-5    & 20186-03-03-02 & 1997 Aug 15        & 07:33:36   \\
\tableline\tableline
 $\beta$-6    & 20402-01-45-03 & 1997 Sep 09        & 06:15:32    \\
\tableline     
 $\beta$-7    & 20402-01-45-03 & 1997 Sep 09        & 08:04:28  \\
\tableline      
 $\beta$-8    & 20402-01-45-03 & 1997 Sep 09        & 09:21:28  \\
\tableline     
 $\beta$-9    & 20402-01-45-03 & 1997 Sep 09        & 09:50:36  \\
\tableline\tableline
 $\alpha$-10  & 50125-01-04-00 & 2002 Jul 27        & 07:30:50    \\
\tableline     
 $\alpha$-11  & 50125-01-04-00 & 2002 Jul 27        & 10:24:48   \\
\tableline\tableline
 $\alpha$-12  & 50125-01-05-00 & 2002 Jul 28        & 06:57:37   \\
\tableline
\enddata
\tablecomments{Observation IDs and dates of the 12 epochs. The Start Time is used as a zero-point in the figures unless stated otherwise. The IDs are based on the \citet{bell00} classifications.}
\label{tbl-1}
\end{deluxetable}

\begin{deluxetable}{ccccccccc}
\tablecaption{{\bf Linear Pearson Correlation Coefficients}}
\tablewidth{0pt}
\tablehead{ \colhead{ID} & \colhead{Group} & \colhead{TF}& \colhead{BBN}& \colhead{BBF}&  \colhead{BBT}&  \colhead{PLN} & \colhead{PLF}& \colhead{PLI}}
\startdata
 $\beta$-1    & 1  &  0.96   & -0.62  &  0.84  &  0.87    &  0.87     & 0.92   &  0.52 \\
\tableline
 $\beta$-2    & 1  &  0.93   & -0.63  &  0.76 &   0.86   &   0.87    &  0.91  &   0.58 \\
\tableline
$\beta$-3    & 1   &  0.96   & -0.62  &  0.84 &   0.90   &   0.81    &  0.88  &   0.39 \\
\tableline 
$\beta$-4    & 1  &   0.96   & -0.66  &  0.89 &   0.89   &   0.88    &  0.94  &   0.44 \\
\tableline
$\beta$-5    & 1  &   0.96   & -0.57  &  0.86 &   0.85   &   0.90    &  0.93  &   0.62 \\
\tableline\tableline 
$\beta$-6    & 2  &   0.90   & -0.53  &  0.89 &   0.88   &   0.79    &  0.78  &   0.53 \\
\tableline 
$\beta$-7    & 2  &   0.76   & -0.42  &  0.78 &   0.75   &   0.63    &  0.59  &   0.56 \\
\tableline
$\beta$-8    & 2  &   0.84   & -0.40  &  0.54 &   0.63   &   0.68    &  0.75  &   0.40 \\
\tableline 
$\beta$-9    & 2  &   0.87   & -0.36  &  0.67 &   0.63   &   0.76    &  0.74  &   0.39 \\
\tableline\tableline  
$\alpha$-10  & 3  &   0.90   & -0.10  &  0.73 &   0.48   &   0.67    &  0.77  &   0.47 \\
\tableline
$\alpha$-11  & 3  &   0.87   & -0.11  &  0.57 &   0.31   &   0.74    &  0.83  &   0.46 \\
\tableline 
$\alpha$-12  & 3  &   0.87   & -0.11  &  0.63 &   0.43   &   0.64    &  0.75  &   0.41 \\
\tableline
\enddata
\tablecomments{Correlation of spectral features to the QPO~frequency. The existence of a correlation is believable for values of $|r|>0.40$ and highly significant for values of $|r|>0.70$. In most cases, the correlation to the power law flux is strongest. The abbreviations are as follows. TF: total flux; BBN: blackbody normalization; BBF: blackbody flux; BBT: blackbody temperature; PLN: power law normalization; PLF: power law flux; PLI: power law index. The Groups are as defined in the text.}
\label{tbl-2}
\end{deluxetable}

\begin{deluxetable}{cccccc}
\tablecaption{{\bf Partial Correlation Coefficients}}
\tablewidth{0pt}
\tablehead{ \colhead{ID} &  \colhead{Group} & \colhead{PLF | BBF}& \colhead{PLF | BBT}& \colhead{BBF | PLF} & \colhead{BBT | PLF}}
\startdata
 $\beta$-1  & 1  &  0.84     &  0.67     &  0.67     &  0.37   \\  
\tableline
 $\beta$-2  & 1  &  0.84     &  0.65     &  0.54     &  0.37    \\ 
\tableline
 $\beta$-3   & 1 &  0.87     &  0.49     &  0.82     &  0.58    \\ 
\tableline
 $\beta$-4  & 1  &  0.82     &  0.74     &  0.66     &  0.48    \\ 
\tableline
 $\beta$-5   & 1 &  0.84     &  0.75     &  0.68     &  0.39    \\ 
\tableline\tableline
 $\beta$-6  & 2  &  0.62     &  0.16     &  0.81     &  0.64    \\ 
\tableline
 $\beta$-7   & 2 &  0.72     &  0.08     &  0.84     &  0.58    \\ 
\tableline
 $\beta$-8   & 2 &  0.80     &  0.58     &  0.64     &  0.31    \\ 
\tableline
 $\beta$-9   & 2 &  0.81     &  0.52     &  0.77     &  0.16    \\ 
\tableline\tableline 
 $\alpha$-10  & 3 & 0.80     &  0.79     &  0.77     &  0.54     \\
\tableline
 $\alpha$-11  & 3 & 0.80     &  0.83     &  0.45     &  0.30     \\
\tableline
 $\alpha$-12  & 3 & 0.81     &  0.78     &  0.72     &  0.53     \\
\tableline
\enddata
\tablecomments{Partial correlation of spectral features to the QPO~frequency. The first column (PLF --- BBF) is the partial of the QPO~frequency and power law flux with the effect of blackbody flux removed. Note that the correlations are strong in most of the cases. The second column (PLF --- BBT) relates the QPO~frequency to power law flux, removing blackbody temperature. These correlations are weaker, but significant in most Group~1 cases. The third column (BBF --- PLF) shows the QPO~frequency correlation to blackbody flux, removing power law flux. Most are believable, suggesting a complex interplay between the power law and blackbody features. The fourth column (BBT --- PLF) shows the QPO~frequency correlation to blackbody temperature, removing power law flux. In this case, nearly all correlations drop below significance showing that power law flux may trace QPO behavior better than blackbody temperature.}
\label{tbl-3}
\end{deluxetable}

\begin{deluxetable}{ccccc}
\tablecaption{{\bf Length of Frequency Dip}}
\tablewidth{0pt}
\tablehead{ \colhead{ID} & \colhead{Group} & \colhead{Min Freq (Hz)}& \colhead{Time at Min (s)}& \colhead{Fraction at Min}}
\startdata
 $\beta$-1    & 1 & 2.2 & 180 & 0.31\\
\tableline
 $\beta$-2    & 1 & 2.2 & 240 & 0.34 \\
\tableline
 $\beta$-3    & 1 & 2.2 & 160 & 0.24 \\
\tableline
 $\beta$-4    & 1 & 2.4 & 160(50) & 0.27 \\
\tableline
 $\beta$-5    & 1 & 2.2 & 180 & 0.34  \\
\tableline\tableline
 $\beta$-6    &  2 &  2.4 & 80(20) & 0.15 \\
\tableline
 $\beta$-7    &  2 &  2.2 & 60 & 0.12 \\
\tableline
 $\beta$-8    & 2 &  2.2 & 100 & 0.15 \\
\tableline
 $\beta$-9    & 2 &  2.7 & 130(0) & 0.21 \\
\tableline\tableline
 $\alpha$-10  & 3 &  2.7 & 560(0) & 0.50 \\
\tableline 
 $\alpha$-11  &  3  &  2.7 & 520(0) & 0.44 \\
\tableline 
 $\alpha$-12  & 3 &  3.1 & 800(0) & 0.63 \\
\tableline
\enddata
\tablecomments{Approximate length of time QPO stays near lowest frequency. The third column shows the minimum QPO~frequency. The fourth column shows the length of time the QPO stays within 0.5~Hz of the minimum frequency. The number in parentheses is the time during which the QPO~frequency is below 2.5~Hz if it is not equal to the listed time. The fifth column shows the approximate fraction of the dip length spent at the minimum frequency. In four cases, the frequency does not drop below 2.5~Hz. In general, Group~2 events spend less time at the minimum frequency than Group~1 events.}
\label{tbl-4}
\end{deluxetable}

\begin{deluxetable}{cccccc}
\tablecaption{{\bf Gaussian Fit Parameters}}
\tablewidth{0pt}
\tablehead{\colhead{ID} & \colhead{Group} & \colhead{$A$}& \colhead{$\sigma_A$}& \colhead{$C_1$}& \colhead{$f_{int}$}}
\startdata
 $\beta$-1 & 1 & 0.70 & 2.20 & 18.24 & 82.10  \\
\tableline	   					    
 $\beta$-2 & 1 & 1.10 & 2.39 &  9.98 & 85.96  \\
\tableline	   					    
 $\beta$-3 & 1 & 0.90 & 2.56 & 10.32 & 94.75  \\ 
\tableline	   					    
 $\beta$-4 & 1 & 1.09 & 1.85 & 12.14 & 89.26  \\
\tableline
 $\beta$-5 & 1 & 0.86 & 2.23 & 16.12 & 91.85  \\
\tableline\tableline
 $\beta$-6 & 2 & 0.80 & 3.64 & -2.56 & 94.49  \\ 
\tableline
 $\beta$-7 & 2 & 0.62 & 3.26 & -1.61 & 92.97  \\ 
\tableline
 $\beta$-8  & 2 & 0.58 & 5.16 & -18.07 & 97.10 \\
\tableline
 $\beta$-9  & 2 & 0.56 & 6.11 & -16.62 & 75.23 \\
\tableline
\enddata				   							      
\tablecomments{Gaussian fit parameters calculated for the ``trigger spike'' in $\beta$-class light-curves. The column labels are as follows: A = the normalized amplitude; $\sigma_A$ = the width (standard deviation) of the gaussian; $C_1$ = the slope of the light-curve background; $f_{int}$ = the integrated count rate. They are grouped by a combination of spike strength, spike width, and the sign of the underlying slope.}
\label{tbl-5}			   
\end{deluxetable}


\clearpage
\begin{thebibliography}{}

\bibitem[Belloni et al.(2000)]{bell00} Belloni~T., Klein-Wolt~M., M\'{e}ndez~M., van~der~Klis~M., \& van~Paradijs~J. 2000, A\&A, 355, 271.
\bibitem[Castro-Tirado et al.(1992)]{cast92} Castro-Tirado~A.~J., Brandt~S., \& Lund~N. 1992 IAUC, 5590, 2.
\bibitem[Eikenberry et al.(1998)]{eiken98} Eikenberry~S.~S., Matthews~K., Morgan~E., Remillard~R.~A., Nelson~R.~W. 1998, ApJ, 494, L61.
\bibitem[Eikenberry et al.(2000)]{eiken00} Eikenberry~S.~S., Matthews~K., Muno~M., Blanco~P.~R., Morgan~E.~H., Remillard~R.~A. 2000, ApJ, 532, L33.
\bibitem[Fender \& Pooley(1998)]{fender98} Fender~R.~P. \& Pooley~G.~G. 1998, MNRAS, 300, 573.
\bibitem[Feroci et al.(1999)]{feroci99} Feroci~M., Matt~G., Pooley~G.~G., Costa~E., Tavani~M., Belloni~T. 1999, A\&A, 351, 985.
\bibitem[Klein-Wolt et al.(2002)]{kleinwolt02} Klein-Wolt~M., Fender~R.~P., Pooley~G.~G., Belloni~T., Migliari~S., Morgan~E.~H., van~der~Klis~M. 2002, MNRAS, 331, 745.
\bibitem[Markwardt et al.(1999)]{markwardt99} Markwardt~C.~B., Swank~J.~H., \& Taam~R.~E. 1999, ApJ, 513, L37.
\bibitem[Mirabel et al.(1998)]{mirabel98} Mirabel~I.~F., Dhawan~V., Chaty~S. Rodr\'\i guez~L.~F., Mart\'\i ~J., Robinson~C.~R., Swank~J., \& Geballe~T.~R. 1998, A\&A, 330, L9.
\bibitem[Mirabel \& Rodriguez(1994)]{mirabel94} Mirabel~I.~F. \& Rodr\'\i guez~L.~F. 1994, Nature, 371, 46.
\bibitem[Moffat, A. F. J. (1969)]{moffat69} Moffat~A.~F.~J. 1969 , A\&A, 3, 455.
\bibitem[Muno et al.(1999)]{muno99} Muno~M.~P., Morgan~E.~H., \& Remillard~R.~A. 1999, ApJ, 527, 321.
\bibitem[Ransom et al.(2002)]{ransom02} Ransom~S.~M., Eikenberry~S.~S., \& Middleditch~J. 2002, ApJ, 124, 1788.
\bibitem[Rothstein et al.(2005)]{droth} Rothstein~D.~M., Eikenberry~S.~S., \& Matthews~K. 2005, ApJ, 626, 991
\end{thebibliography}
\end{document}